# Fast Fraction-Integer Method for Computing Multiplicative Inverse

Hani M. AL-Matari[1] and Sattar J. Aboud[2] and Nidal F. Shilbayeh[1]
[1] Middle East University for Graduate Studies, Faculty of IT, Jordan-Amman
[2] Information Technology Advisor, Iraqi Council of Representatives, Baghdad-Iraq

**Abstract -** Multiplicative inverse is a crucial operation in public key cryptography, and been widely used in cryptography. Public key cryptography has given rise to such a need, in which we need to generate a related public and private pair of numbers, each of which is the inverse of the other. The basic method to find multiplicative inverses is Extended-Euclidean method. In this paper we will propose a new algorithm for computing the inverse, based on continues subtract fraction from integer and divide by fraction to obtain integer that will be used to compute the inverse d. The authors claim that the proposed method more efficient and faster than the existed methods.

## 1. Introduction

Modular arithmetic plays an important role in cryptography. Many public-key schemes [2] involve modular exponentiation. Modular inversion, the computation of $b^{-1} \bmod a$ has a part in exponentiation based on addition-subtraction chains [6], as well as other applications in such public key systems.

The multiplicative inverse of $e$ modulus $n$ is an integer $d$ such that $e*d \equiv 1 \bmod n$, $d$ is called the inverse of $e$ and denoted $e^{-1}$ [5]. The study of inverse calculation was an intractable science due to lack of real improvement, the modulus inverse problem is a lot more difficult to solve [1]. However, there were only a few methods.

The first one is trivial and lengthy in calculating the inverse, because it is a sequential search. It starts by $d = 1$, keep on adding 1 to $d$ until $e*d \equiv 1 \bmod n$.

In [3] Euclidian described the algorithm in his book, Elements, written around 300 B.C. It is the oldest nontrivial algorithm that has survived to the present day, and it is still a good one. Euclid's algorithm is an efficient method to calculate the greatest common divisor of two integers without factoring them.

Euclidian algorithm can also compute the inverse of a number modulo $n$, sometimes this is called the extended Euclidean algorithm, this method is based on the idea that if $n > a$ then $\gcd(a,n) = \gcd(a, n \bmod a)$, also on finding $a*x + y*n = 1$ in which $x$ is the multiplicative inverse.

Euclidian algorithm is approximately irrelevant to $e$ or $n$, but other algorithms are affected by $e$ and the modulus $n$.

## 2. Previous methods

In this section we will describe the methods that deal with the computing multiplicative inverse which are as follows:

### 2.1. Euclid algorithm

This method is based on the idea that if $n > e$ then $\gcd(e,n) = 1$, also on finding $e*x + y*n = 1$ in which $x$ is the multiplicative inverse of $e$ [4]. The algorithm is iterative and can be slow for large numbers. Knuth showed that the average number of divisions performed by the algorithm is $0.843*\log_2(n) + 1.47$ [2].

The method needs 8 variables, and used subtraction, multiplication, division, and comparison as operations, the complexity of $O(\log n)$.

**Algorithm**

Input: $e \in Z_n$ such that $\gcd(e,n) = 1$.

Output: $e^{-1} \bmod n$ where $e^{-1} = i$ provided that it exists.

The algorithm is as follows:
1. Set $g \leftarrow n; u \leftarrow e; i \leftarrow 0; v \leftarrow 1$;
2. While $u > 0$ do the following:
   $q \leftarrow \lfloor g/u \rfloor; t \leftarrow g - q*u$;
   $g \leftarrow u; u \leftarrow t; t \leftarrow i - q*v$;
   $i \leftarrow v; v \leftarrow t$;
3. If $i < 0$ then
   $i \leftarrow n + i$;
4. $e^{-1} \leftarrow i$

**Example**

Let $e \leftarrow 7; n \leftarrow 60$

| g  | u | i   | v   | q | t   |
|----|---|-----|-----|---|-----|
| 60 | 7 | 0   | 1   | 0 | 0   |
| 7  | 4 | 1   | -8  | 8 | -8  |
| 4  | 3 | -8  | 9   | 1 | 9   |
| 3  | 1 | 9   | -17 | 1 | -17 |
| 1  | 0 | -17 | -52 | 3 | -52 |

$e^{-1} \leftarrow n + i = 60 + (-17) = 43$

**2.2. Stein Method**

In 1967, Stein introduced an inverse algorithm [7] and later improved by Penk Knuth. It is based on the observation that $\gcd(x,y) = \gcd(x/2, y)$ if $x$ is even, also $\gcd(x,y) = 2$, $\gcd(x/2, y/2)$ if both $x, y$ are even, and $\gcd(x,y) = \gcd((x-y)/2, y)$ if $x, y$ are both odd.

The algorithm needs about 11 variables, and uses addition, subtraction, multiplication, division and comparison, the complexity is $O(\log n)$.

**Algorithm**

Input: $e \in Z_n$ such that $\gcd(e,n) = 1$.

Output: $e^{-1} \bmod n$ provided that it exists.
The algorithm is as follows:
While $e$ and $n$ is even do
  $e \leftarrow \lfloor e/2 \rfloor; n \leftarrow \lfloor n/2 \rfloor$;
$u_1 \leftarrow 1; u_2 \leftarrow 0; u_3 \leftarrow e; v_1 \leftarrow n; v_2 \leftarrow 1 - e; v_3 \leftarrow n$; If $e$ is odd then
  $t_1 \leftarrow 0; t_2 \leftarrow -1; t_3 \leftarrow -n$;
Else $t_1 \leftarrow 1; t_2 \leftarrow 0; t_3 \leftarrow e$;
Repeat
  While $t_3$ is even do
    $t_3 \leftarrow \lfloor t_3/2 \rfloor$
  If $t_1$ and $t_2$ is even then
    $t_1 \leftarrow \lfloor t_1/2 \rfloor; t_2 \leftarrow \lfloor t_2/2 \rfloor$;
  Else $t_1 \leftarrow \lfloor (t_1+n)/2 \rfloor; t_2 \leftarrow \lfloor (t_2-e)/2 \rfloor$;
  If $(t_3 > 0)$ then
    $u_1 \leftarrow t_1; u_2 \leftarrow t_2; u_3 \leftarrow t_3$;
  Else $v_1 \leftarrow n - t_1; v_2 \leftarrow -(e + t_2); v_3 \leftarrow -t_3$;
    $t_1 \leftarrow u_1 - v_1; t_2 \leftarrow u_2 - v_2; t_3 \leftarrow u_3 - v_3$;
  If $(t_1 < 0)$ then
    $t_1 \leftarrow t_1 + n; t_2 \leftarrow t_2 - e$;
Until $t_3 = 0$;

$e^{-1} \leftarrow u_1$;

**Example**

Let $e = 7; n = 60$;

| e | n  | $u_1$ | $u_2$ | $u_3$ | $v_1$ | $v_2$ | $v_3$ | $t_1$ | $t_2$ | $t_3$ |
|---|----|-------|-------|-------|-------|-------|-------|-------|-------|-------|
| 7 | 60 | 1     | 0     | 7     | 60    | -6    | 60    | 0     | -1    | -60   |
|   |    |       |       |       |       |       |       | 30    | -4    | -30   |
|   |    |       |       |       |       |       |       | 15    | -2    | -15   |
|   |    |       |       |       | 45    | -5    | 15    |       |       |       |
|   |    |       |       |       |       |       |       | -44   | 5     | -8    |
|   |    |       |       |       |       |       |       | 16    | -2    |       |
|   |    |       |       |       |       |       |       | 8     | -1    | -4    |
|   |    |       |       |       |       |       |       | 34    | -4    | -2    |
|   |    |       |       |       |       |       |       | 17    | -2    | -1    |
|   |    |       |       |       | 43    | 1     |       |       |       |       |
|   |    |       |       |       |       |       |       | -42   | 5     | 6     |
|   |    |       |       |       |       |       |       | -18   | -2    |       |
|   |    |       |       |       |       |       |       | 9     | -1    | 3     |
|   |    |       |       |       | 9     | -1    | 3     |       |       |       |
|   |    |       |       |       |       |       |       | -43   | 4     | 2     |
|   |    |       |       |       |       |       |       | -26   | -3    |       |
|   |    |       |       |       |       |       |       | 43    | -5    | 1     |
|   |    |       |       |       | 43    | -5    | 1     |       |       |       |
|   |    |       |       |       |       |       |       | 0     | 0     | 0     |

$e^{-1} \leftarrow u_1 = 43$

**2.3. Gordon Method**

In 1989, Gordon [2] described another algorithm for computing an inverse. It is based on the observation that $q$ at Euclidian method does not need to be the remainder of $n/a$ but it can be any power of 2 up to that limit [4]. The algorithm needs about 9 variables, and uses addition, subtraction, comparison, and shifting. The complexity of the algorithm is $O(\log n)$

**Algorithm**

Input: $e \in Z_n$ such that $\gcd(e,n) = 1$.

Output: $e^{-1} \bmod n$ provided that it exists.
The algorithm is as follows:
$g \leftarrow n; i \leftarrow 0; v \leftarrow 1; u \leftarrow e$;
Repeat
  $s \leftarrow -1; p \leftarrow 0$;
  If $u > g$ then

$t \leftarrow 0;$
Else
  $p \leftarrow 1; t \leftarrow u;$
  While $(t \leq g)$ do
      $s \leftarrow s + 1;$
      $t \leftarrow$ Left shift $t$ by 1;
   $t \leftarrow$ Right shift $t$ by 1;
  $t \leftarrow g - t; g \leftarrow u; u \leftarrow t; t \leftarrow i; i \leftarrow v;$
  If $p = 1$ then
      $v \leftarrow$ Left shift $v$ by $s;$
      $t \leftarrow t - v;$
      $v \leftarrow t;$
  Until $u = 0$ or $u = g;$
  If $i < 0$ then
      $i \leftarrow n + i;$
$e^{-1} \leftarrow i;$

**Example**
Let $e \leftarrow 7; n \leftarrow 60$

| g  | u | i  | v   | s | p | t   |
|----|---|----|-----|---|---|-----|
| 60 | 7 | 0  | 1   | 0 | 1 | 14  |
|    |   |    |     | 1 |   | 28  |
|    |   |    |     | 2 |   | 58  |
|    |   |    |     | 3 |   | 112 |
|    |   |    |     |   |   | 56  |
| 7  | 4 |    |     |   |   | 4   |
|    |   | 1  |     |   |   | 0   |
|    |   |    | 8   |   |   | −8  |
|    |   |    |     |   | −1 | 0  |
|    |   |    |     |   | 1 | 4   |
|    |   |    |     |   | 0 | 8   |
| 4  | 3 |    |     |   |   | 3   |
|    |   | -8 |     |   |   | 1   |
|    |   |    |     |   |   | 9   |
|    |   |    | 9   |   |   |     |
|    |   |    |     |   | −1 | 0 |
|    |   |    |     |   | 1 | 3   |
|    |   |    |     |   | 0 | 6   |
|    |   |    |     |   |   | 3   |
| 1  | 1 |    |     |   |   | 1   |
|    |   |    | 9   |   |   | −8  |
|    |   |    |     |   |   | −17 |
|    |   |    | −17 |   | −1 | 0  |
|    |   |    |     |   | 1 | 1   |
|    |   |    |     |   | 0 | 2   |
|    |   |    |     |   | 1 | 4   |
|    |   |    |     |   |   | 2   |
| 1  |   |    |     |   |   | 1   |
|    |   | −17|     |   |   | 9   |
|    |   |    | −3  |   |   | 43  |
|    |   |    | 43  |   |   |     |

$e^{-1} \leftarrow 60 - 17 = 43$

**2.4. Baghdad algorithm**

In 2004, Sattar Aboud [6] introduced another algorithm entitled "Baghdad method" to calculate the inverse. The idea behind Baghdad method is very simple involving adding 1 to the modulus $n$ and then divides the result by the exponent $e$. Then keep on adding the result to the modulus $n$ and divide the new result by the exponent $e$ until an integer is obtain.

The algorithm needs only 5 variables, and uses addition and division only. The complexity of the algorithm is $O(\log n)$

**Algorithm**
Input: $e \in Z_n$ such that $\gcd(e, n) = 1$

Output: $e^{-1} \bmod n$ provided that it exists
The algorithm is as follows:
Set $d \leftarrow 1;$
Repeat
   $d = (d + n)/e;$
Until $d$ is integer
$e^{-1} \leftarrow d;$

**Example**
Let $e \leftarrow 7; n \leftarrow 60;$

| d | result |
|---|--------|
| (1+60)/7 | not integer |
| (61+60)/7 | not integer |
| (121+60)/7 | not integer |
| (181+60)/7 | not integer |
| (241+60)/7 | integer match |

$e^{-1} \leftarrow d = 43$

## 3. Fast Fraction-Integer Method

The idea behind the proposal method is a very simple, based on continues subtract fraction from integer and divide by fraction to obtain integer that will be used to compute the inverse $d$. The algorithm needs only 6 variables, and uses addition and division only. The complexity of the algorithm is $O(\log n)$

**Algorithm**
Input: $e \in Z_n$ such that $\gcd(e, n) = 1$

Output: $e^{-1} \bmod n$ provided that it exists.
The algorithm as follows:
$r : real;$
$i = 1;$
$s_f = (n + 1 \bmod e)/e;$
$d_f = (n \bmod e)/e;$
If $s_f = 0$ then
   Stop;
Else
  Repeat
      $r = ((i - s_f)/d_f);$

$i = i + 1;$
Until $r$ is integer
$d = (n*(r+1)) + 1/e;$

**Example**
Let $e \leftarrow 7; n \leftarrow 60;$

| $i$ | $s_f$ | $d_f$ | $r$ |
|---|---|---|---|
| 1 | 0.71428 | 0.57142 | 0.50001 |
| 2 | | | 2.25004 |
| 3 | | | 4.00000 |

$d = (60*(r+1)) + 1/e$
$= (60*(4+1)) + 1/7$
$= (60*5) + 1/7$
$= 301/7$
$= 43$

**3.1 Proof of Fast Fraction-Integer Method**
In order to prove the algorithm, we need to prove that the algorithm will give integer number only when $d$ is the inverse of $e$. As we know that if $d$ is the inverse of $e$ then
1. Both $e$, $d$ are positive integer numbers between $[1..n]$ ……. …………………… (1)
2. $\gcd(e,n) = 1$ ……………..……..……... (2)
3. $e*d \equiv 1 \mod n$, it means that $e*d = 1 + k*n$ for $k \in Z$ …………………….....…... (3)

So $d = (1 + k*n)/e$
$= 1/e + k*n/e$ …...………………….. (4)

From the algorithm of Fast Fraction-Integer Method we see that $d = (n*(r+1)) + 1/e$; this will repeated $i$ times until $d$ ………....... (5)

From that we know that the algorithm above is correct for $i = k$, but if this is the case we need to prove that (5) will give none integer for all values of $i < k$, and the only integer value is when $i = k$, so we know $d$ is an integer so $(1 + k*n)/e$ is also integer for integer value of $k$.

Then we need to proof that $(1 + i*n)/e$ is never an integer for all values of $i$ between $[1, k-1]$. Assume that there is another value of $i$ where $i < i < k$ such that $d = (1 + i*n)/e$ is also an integer, it means that $i = k - 1$ ------------ (6)

Then $d = (1 + (k-1)*n)/e$ will be integer. So
$d = (1 + k*n - n)/e$
$= (1 + k*n)/e - n/e$
$= 1/e + k*n/e - n/e$

But by $1/e + k*n/e$ is integer, and by that $\gcd(e,n)$ should be 1. So if there is no greater common divisor between $e$ and $n$ except 1, that means $n/e$ is a non integer value.

Thus subtracting a non integer value form an integer value will yield $d$ is not an integer. This will contradict our assumption (that $d$ is an integer).

Now assume that there exist $i = k - q$ such that $d$ is an integer for $q$ between $[1, k-1]$. Then
$d = (1 + (k-q)*n)/e$
$= 1/e + k*n/e - q*n/e$

If this to be integer then $q*n/e$ must be integer, but since $\gcd(e,n) = 1$ then $q$ must be a multiple of $e$ so $d = 1/e + k*n/e - x*n$ (5)

This will lead to $d$ being a negative number $d < 0$ but from definition we know that both $e$ and $d$ must be positive (1) so there is no values for $x$ that satisfy the definition. So the only value for $q$ that satisfy the conditions is when $q = 0$ and that $i = k$.

**3.2 Problem of Fast Fraction-Integer method**
We have proved that Fast Fraction-Integer algorithm is correct, but the question is that is it implemental? Yes the algorithm will terminate giving the correct answer when implemented using the computer programming languages.

Let $dm$ be the mathematical value of $d$ where $d = dm$. Let $dc$ be the calculated value of $d$ in the computer memory and registers. Let $\xi$ be the error in calculating, between the mathematical value and the computer value (round off error).
So $dm = (1m + km*nm)/em$
$= 1m/em + km*nm/em$
$= (1/e)m + (k*n/e)m$

But we know that the calculated value of fractions is never exactly as the mathematical value for big values of $e$ that when used to divide 1 and $n$ will give a cyclic fraction number.

So $(1/e)m = (1/e)c + \xi 1$ and $(n/e)m = (n/e)c + \xi 2$ where $\xi 1 << (1/e)c$, $\xi 2 << (n/e)c$ and $dc = (1/e)c + \xi 1 + k*\xi 2$ such errors will yield that either $dm \leq dc$ or $dm \geq dc$, $dm - dc$ if and only if $\xi 1 + k*\xi 2 = 0$ it means that $(1/e)m = (1/e)c$, $(n/e)m = (n/e)c$ We know that the error $\xi 1, \xi 2$ is small, but multiplying $\xi 2$ with $k$ will give big value to the error and the error will multiply by $k$, so as $k$ is increasing the error also will increase so the best approach is to use small values for $e$.

## 4. Conclusions

For security reasons, cryptography recommends smaller values for public keys and bigger values for private keys [4]. The suggested algorithm needs lower values for public keys (lower value of $e$) and higher values for private key, which is fully compatible with the preferred cryptography algorithm. The method is simple, fast and needs less storage, and its complexity is also less.